\documentclass[prc,aps,showpacs,nofootinbib,twocolumn]{revtex4}
\usepackage{epsfig}
\begin{document}
\title{The nuclear moment of inertia and spin distribution of nuclear levels}
\author{Y. Alhassid,$^{1}$  G.F. Bertsch,$^{2}$ L. Fang,$^{1}$ and S. Liu$^{1}$}
\affiliation{$^{1}$Center for Theoretical Physics, Sloane Physics
Laboratory, Yale University, New Haven, CT 06520\\
$^{2}$Department of Physics and Institute of Nuclear Theory,
Box 351560\\
University of Washington
Seattle, WA 98915}
\def\be{\begin{equation}}
\def\ee{\end{equation}}
\def\sumk{\sum_k}
\def\ad{a^\dagger_k}
\def\adb{a^\dagger_{\bar k}}
\def\a{a_k}
\def\ab{a_{\bar k}}
\def\Tr{{\rm Tr}}
\def\lb{\langle}
\def\rb{\rangle}
\begin{abstract}
  We introduce a simple model to calculate the nuclear moment of inertia at
finite temperature. This moment of inertia describes the spin distribution
of nuclear levels in the framework of the spin-cutoff model.  Our model
is based on a deformed single-particle Hamiltonian with pairing
interaction and takes into account fluctuations in the pairing gap. We derive
a formula for the moment of inertia at finite temperature that generalizes
the Belyaev formula for zero temperature. We show that a number-parity
projection explains the strong odd-even effects observed in shell model
Monte Carlo studies of the nuclear moment of inertia in the iron region.
\end{abstract}

\maketitle

\section{Introduction}

 The shell model Monte Carlo (SMMC) method has proven to be quite accurate for
calculating the nuclear level density in the range of excitation energies up to
several tens of MeV~\cite{na97,al99,al00}. The advantage of the SMMC method
is that it can be used
to calculate thermal observables in model spaces that are orders of magnitude
larger than can be treated by conventional
diagonalization methods.  In practice, most of the SMMC calculations are
carried out in
 truncated spaces (e.g., one major shell), hence the limitation
on the excitation energy. Correlations become less important at higher
temperatures, and the results of the truncated
SMMC calculations can be extended to higher temperatures or excitation
energies by taking into account the effects of a larger space in the
independent-particle model~\cite{al03}.

   The SMMC method can also be used to calculate the distribution of
nuclear spins
at finite temperature~\cite{spin-proj}.  However, this requires
spin projection, and the associated computational effort is rather large.
For general purposes such as constructing tables for large numbers of
nuclei, simplified models are thus invaluable. The aim of this paper is
to construct
 and study such a simple model that can reproduce well the spin
distributions of the
microscopic SMMC method.

  A common assumption~\cite{il92,ra97} in global parameterizations of
nuclear level
densities is that the spin distribution follows the spin-cutoff model
\begin{equation}\label{spin-cutoff}
{\rho_J \over \rho} = {2 J + 1\over 2 \sqrt{2\pi}\sigma^3}
e^{-{J(J+1)\over 2 \sigma^2 }}
 \;.
\end{equation}
Here $\rho$ is the total level density counting all $M$ states in a
$J$ multiplet, while $\rho_J$ is the density of spin-$J$ levels without
the $2J+1$ degeneracy factor. Thus $\sum_J (2J+1) \rho_J = \rho$.
The parameter $\sigma$ is known as
the spin cutoff parameter.  The quantities $\rho,
\rho_J,$ and $\sigma$ are all functions of excitation energy.

The model can be derived assuming that the individual nucleon spins add up as
independent random vectors~\cite{er60}, $\vec J= \sum_i \vec j_i$, leading to a
Gaussian distribution of the spin vector $P(\vec J) \propto e^{-\vec
J^2/2\sigma^2}$.
Integrating over the orientation of $\vec J$, we have
$P(J) \propto J^2 e^{-J^2/2\sigma^2}$, where J is the magnitude of the
angular momentum  (the pre-exponential factor $J^2$ comes from the
Jacobian in spherical coordinates).
We recover Eq.~(\ref{spin-cutoff}) by making the semiclassical
substitution $J \to J+1/2$; the spin cutoff parameter is then given by
\begin{equation}\label{sigma}
\sigma^2=\langle J_z^2\rangle = {1\over 3}\langle \vec J^2\rangle  \;.
\end{equation}
In thermal ensembles it is common to define an effective moment of inertia
$I$ by the relation between $\langle \vec J^2\rangle$ and temperature $T$,
which we can write as
\begin{equation}\label{sigma-I}
I = {\hbar^2 \over T } \sigma^2 \;.
\end{equation}
In many of the empirical parameterizations,  $\sigma$ is determined by
this formula using for $I$ the rigid-body moment of inertia, $I= 2 m A (r_0 A^{1/3})^2/5$, where
$r_0 \approx 1.2-1.3$ fm is the usual
nuclear radius parameter, $A$ is the mass number and $m$ is the nucleon mass.
Other treatments of $\sigma$ based on the independent-particle model
have also been proposed~\cite{go96}.
SMMC calculations of nuclei in the $A \sim 50 - 70$ mass region
show that the assumption of a rigid-body moment of inertia breaks
down at low excitation energies starting somewhat
below the neutron separation energy,  especially in even-even nuclei.
The effect has a clear odd-even mass
dependence.  Furthermore, at the lowest excitations, deviations are
 observed from the spin-cutoff model itself, and odd-even staggering
effects (in spin) can be seen.
Here we will show that
a fairly simple model based on a fixed deformation and a fluctuating
pairing field reproduces very well the detailed SMMC results for the effective
moment of inertia at finite temperature. In particular, odd-even effects
observed in the microscopic SMMC calculations are nicely reproduced by a
number-parity projection
method~\cite{go81,ro98,ba99,fl01}. We would therefore advocate this model
for global calculations of the spin distributions below the neutron
separation energy. Such distributions are needed for theoretical
estimates of nucleosynthesis reaction rates~\cite{ra97},
among other applications.

   Our model is based on the static path approximation~\cite{mu72,al84,la88}
to the BCS Hamiltonian~\cite{bcs}.  BCS theory is valid in the limit
 when the mean level spacing is much smaller than the pairing gap.
  However, this condition
does not hold in the finite nucleus, in which case fluctuations
must be taken into
account. A similar situation occurs in ultra-small metallic particles
whose linear size is smaller than $\sim 3$ nm~\cite{nano01}. Theoretical
studies have indicated that pairing correlations in the crossover from
BCS to the fluctuation-dominated regime are manifested through their
number-parity dependence.  Odd-even effects that originate in pairing
 correlations were found in the SMMC heat capacity of nuclei~\cite{li01}.
 Such effects were also observed in the heat capacities of rare-earth nuclei
that were extracted from level density measurements~\cite{sc01,gu03}.
Finite-temperature pairing correlations at a fixed number of particles
were also studied in Ref.~\cite{fr03}.

In this paper, we first discuss in Section \ref{sec:formal}
general aspects of calculating the thermal moment of inertia
and projection on number parity.  We work in a grand
canonical ensemble, but the odd-even effects can be extracted
by the number-parity projection operator.  In Section \ref{sec:model},
we apply the formalism to a
model Hamiltonian that includes a deformed single-particle field and
a pairing interaction treated in the static path approximation. This
yields a formula for the moment of inertia that is a generalization of the
Belyaev formula~\cite{be61} for zero temperature, explaining the suppression
of the inertia at low temperature.
In Section \ref{sec:number-parity}, we further generalize
the moment-of-inertia formula to take into account odd-even differences,
 making use of the number-parity projection operator.  In Section
\ref{sec:results}, we apply the model to nuclei in the iron region using
the $pfg_{9/2}$ shell with single-particle energies and wave functions
determined from a deformed Woods-Saxon potential. The calculated moments of
inertia are found to be in good agreement with the SMMC calculations.

\section{Formal aspects}\label{sec:formal}

 In general, the SMMC method~\cite{la93,al94} can be used to calculate
thermal expectation values of observables ${\cal O}$
\be\label{observable}
\langle {\cal O}\rangle = {\Tr \left({\cal O} e^{-\beta H}\right)\over Z} \;,
\ee
where
\begin{equation}\label{partition}
Z = \Tr e^{-\beta H}
\end{equation}
 is the nuclear partition function.  $H$ is the nuclear Hamiltonian,
containing rotational invariant one-body and two-body
terms.  In Ref.~\cite{la93,al94} exact particle-number projection was
performed  to calculate the traces
in Eq.~(\ref{observable}) at fixed number of protons and neutrons.

 In Ref.~\cite{spin-proj}, the spin distribution was calculated using spin
projection techniques. For temperatures that are not too low, it was found
that the spin-cutoff model (\ref{spin-cutoff}) describes rather well
the spin distribution but with an energy-dependent moment of inertia.
 The purpose of the present work is to understand the temperature dependence of
the moment of inertia in terms of a simple model. We note, however, that
at the lowest temperatures the SMMC calculations reveal
deviations from the spin-cutoff model (\ref{spin-cutoff}), which are beyond the
scope of the model discussed here.

\subsection{Moment of inertia}\label{sec:inertia}

  In this work we shall assume that the spin distribution can be described by
 Eq.~(\ref{spin-cutoff}), and we therefore only need to calculate the
 variance $\sigma^2$.  The obvious way to do this is to evaluate the
expectation value of the operator ${\cal O} = J^2$ directly from
(\ref{observable}), as is done in SMMC.  However, our model in Section
\ref{sec:model} is based on a deformed Hamiltonian $H_{\rm def}$,
and for such Hamiltonians it is useful to define a moment of inertia
tensor $I_{ij}$ as the response of the nucleus to a rotational
field $\vec\omega$.

 We shall work in the grand-canonical ensemble, replacing $H$ by
$H'=H_{\rm def}-\mu \hat N$ in Eqs.~(\ref{partition}) and (\ref{observable}).
 In the presence of a rotational field, the Hamiltonian
is given by $H'-\vec\omega \cdot \vec J$ and its free energy is
\be\label{free}
 F(\beta,\vec\omega)= - \beta^{-1} \ln {\rm Tr}
e^{-\beta(H'-\vec\omega\cdot \vec J)} \;.
\ee
 The moment of inertia $I_{ij}$ is defined by the expansion of $F$ to
 second order in $\vec \omega$, $F(\beta,\vec\omega) = F(\beta,\omega=0) -
 {1 \over 2} I_{ij}\omega_i\omega_j$, where $\omega_i$ are the components
of $\vec\omega$. Equivalently~\cite{al87}
\begin{equation}\label{inertia}
 I_{ij}= - {\partial^2 F \over \partial \omega_i \partial
\omega_j}\vert_{\omega=0}=\int_0^\beta d\tau \langle J_i(\tau)J_j(0)\rangle \;,
\end{equation}
where
\be\label{inertia-tensor}
 \langle J_i(\tau)J_j(0)\rangle={{\rm Tr} [e^{-\beta H'}(e^{\tau H'} J_i
e^{-\tau H'})J_j] \over Z}
\ee
is the spin response function in imaginary time.
 For a rotationally invariant Hamiltonian, $J_i(\tau)=J_i$, and $I_{ij}=I
 \delta_{ij}$ with $I=\beta \langle J_z^2\rangle = \beta \langle \vec
J^2\rangle/3$, in agreement with Eqs.~(\ref{sigma}) and (\ref{sigma-I}).
 Choosing the cranking axis along the fixed $z$-axis of the laboratory
frame, we can calculate $I$ from \be\label{inertia1} I =
{\partial^2  F \over \partial \omega^2}\vert_{\omega=0}\;, \ee
where\footnote{Here we use a different notation for the angular
  momentum in the laboratory frame to distinguish it from the angular
  momentum in the intrinsic frame.}
\be\label{free1} F(\beta,\omega) = -T \ln {\rm Tr} e^{-\beta
(H'-\omega L_z)} \;, \ee
 and the angular momentum component along $z$ is denoted by
 $L_z$.

 A non-rotational invariant effective Hamiltonian arises in the  mean-field
 approximation when the single-particle potential is deformed. In such a
 case $H_{\rm def}$ describes the Hamiltonian in the intrinsic frame
 of the nucleus.
 The quantity $I_{ij}$ in Eq.~(\ref{inertia}) is then the moment of
  inertia tensor in
 this intrinsic frame, where $J_i$ are the {\it intrinsic} components of
 the angular momentum $\vec J$.  To recover the moment of inertia $I$ in
 (\ref{sigma-I}), it is necessary to integrate over all orientations
 of the intrinsic frame and then use (\ref{inertia1}).
  One obtains the result  (see Appendix B)
\be\label{inertia3}
I={1\over 3}(I_{xx}+ I_{yy}+ I_{zz}) \;.
\ee
 Eq.~(\ref{inertia3}) expresses the effective moment of inertia $I$ in
terms of the intrinsic principal moments $I_{ii}$.

\subsection{Number-parity projection}\label{sec:NP}

 The calculations in the previous Section \ref{sec:inertia} were described
in the grand-canonical ensemble. While this allows the average number of
particles to be specified, it is not precise enough to reproduce odd-even
effects. We note
that the behavior of odd and even nuclei at low temperatures is quite
different;  the spin goes to zero for even nuclei due to pairing, but
remains finite at zero temperature for odd nuclei.  Exact particle-number
projection can be done using the projection operator
 $ P_N = \int_0^{2\pi} e^{i\phi(N-\hat N)} d\phi/2\pi$ as in the SMMC,
  but leads to cumbersome expressions. In
 order to capture the main odd-even effects, it is often sufficient to use
a number-parity projection~\cite{go81,ro98,ba99,fl01} that
distinguishes only between even and odd
number of particles.  The number-parity
projection is defined by
 \be\label{NP-projector}
P_\eta = {1\over 2 } \left( 1 + \eta e^{i\pi \hat N}\right) \;,
\ee
 where $\eta=1$ or -1 describes the projection on an even or odd
number of particles, respectively. Thus, the number-parity
projected partition function is
\be\label{projected-partition}
 Z_\eta = {\rm Tr\,} \left(P_\eta e^{-\beta H'}\right) = {1\over 2} Z
\left[1 + \eta
\langle e^{i\pi \hat N}\rangle \right] \;,
\ee
where the bracket denotes a thermal trace, $\lb {\cal O} \rb \equiv
{\rm Tr\,} \left({\cal O} e^{-\beta H'}\right)/ {\rm Tr\,}e^{-\beta H'}$.
 We can also calculate number-parity projected expectation values of
 observables
\be
\langle {\cal O}\rangle_\eta \equiv {  {\rm Tr\,} \left({\cal O} P_\eta
e^{-\beta H'}\right) \over  {\rm Tr\,} \left(P_\eta e^{-\beta H'}\right)}\;.
\ee
Using (\ref{NP-projector}), we find
\be\label{projected-observable}
\langle {\cal O}\rangle_\eta= {\langle{\cal O}\rangle +\eta
\langle e^{i\pi \hat N}\rangle \langle {\cal O}\rangle_\pi \over
1 +\eta
\langle e^{i\pi \hat N}\rangle} \;,
\ee
where we have used the notation
\be\label{observable-pi}
\langle {\cal O}\rangle_\pi \equiv
 {{\rm Tr\,} \left({\cal O} e^{i\pi \hat N} e^{-\beta H'}\right) \over
{\rm Tr\,} \left(e^{i\pi \hat N} e^{-\beta H'}\right)} \;.
\ee

  The number-parity projected moment of inertia $I_{ij}^\eta$ is defined
from the second-order expansion (in $\vec\omega$) of the number-parity
projected free energy
\be
 F_\eta = - T \ln {\rm Tr}\, \left[P_\eta e^{-\beta (H'- \vec\omega \cdot
\vec J)}\right] \;.
\ee
We find
\begin{equation}
I_{ij}^\eta={\int_0^\beta d\tau\;\langle J_i(\tau)J_j(0)\rangle
+ \eta \langle e^{i\pi\hat N}\rangle
\int_0^\beta d\tau\;\langle J_i(\tau)J_j(0)\rangle_\pi
\over 1+\eta  \langle e^{i\pi\hat N}\rangle} \;,
\end{equation}
where $\langle J_i(\tau)J_j(0)\rangle_\pi$ is defined as in
(\ref{observable-pi}).

\section{Model}\label{sec:model}

We now ask, starting from the independent-particle shell model,
what is the minimal model that will include the most relevant
interaction effects
for calculating the spin distribution. Clearly, the most important
correlations are those associated with the quadrupole deformation and the
pairing field.  Both of these
can be treated in a mean-field approximation, but the mean-field equations
predict sharp transitions that are not supported by more detailed theories.
Thus we go one step further in the finite-temperature theory, using
the static path approximation (SPA)~\cite{mu72,al84,la88} of the
partition function to include time-independent fluctuations of the order
parameters.

 We consider an Hamiltonian composed of an axially deformed
 Woods-Saxon well for the single-particle potential and
orbital-independent pairing for the interaction.  We denote by $\vert
k\rangle$ the single-particle eigenstates in the deformed potential with
energies $\epsilon_k$. They can be divided into degenerate time-reversed
pairs $(k,\bar k)$. For an axially symmetric potential,
$\vert k\rangle=\vert q,\;\mu\rangle$ where $\mu$ is the projection
of the angular momentum on the symmetry axis and $q$ are other labels of
the states. The time-reversed states are defined by
$\vert \bar{k}\rangle=\vert q,\;-\mu\rangle=\vert -k\rangle$ (known as
the BCS phase convention), and we adopt the convention
$k>0\Leftrightarrow \mu>0$.
The Hamiltonian may then be expressed in the form
\be\label{Hamiltonian}
H_{\rm def} = \sum_{k>0} \epsilon_k(\ad\a+\adb\ab) - G P^\dagger P \;,
\ee
where
$P^\dagger$ is the pair creation operator,
$P^\dagger = \sum_{k>0}\ad \adb$, and $G$ is the pairing strength.

\subsection{Static path approximation}\label{sec:SPA}

 The Hamiltonian (\ref{Hamiltonian}) contains a pairing interaction.
Using the Hubbard-Stratonovich transformation, the imaginary-time
propagator $e^{-\beta H'}$
can be written as a functional integral over pairing fields of
propagators that describe non-interacting quasi-particles.  Here we shall
use the SPA, which takes into account only time-independent pairing
fields. The functional integral then reduces to an ordinary integral over
a complex pairing field $\xi$~\cite{mu72}
\be\label{SPA}
e^{-\beta H'} \approx {\beta G\over 2\pi} \int d \xi d\xi^* e^{-\beta G |\xi|^2}
e^{-\beta \sum_{k>0} H_k} \;,
\ee
where
\be \label{pairing-k}
\!\!\! H_k= (\epsilon_k-\mu-G/2) (\ad\a+\adb\ab)-G
\xi^* \ab\a - G \xi \ad\adb + G/2 \;.
\ee

 Our model (\ref{Hamiltonian}) describes nucleons moving in a deformed
well, but it could have been derived from a rotationally invariant
Hamiltonian that included quadrupolar two-body interaction.  This would
introduce five additional integration variables in the SPA integral
 (\ref{SPA}), two representing the intrinsic deformation and three
 representing the orientation of the deformed field~\cite{la88}. This
 integration over the Euler angles of the intrinsic frame is equivalent to
the symmetry restoration described by (\ref{restore}).

\subsubsection{Partition function}\label{sec:partition}

  Using (\ref{SPA}), we can represent the grand-canonical partition
function in the form
\be\label{SPA-Z}
Z={\beta G\over 2\pi} \int d \xi d\xi^* e^{-\beta G |\xi|^2}
\left(\prod_{k>0}{\rm Tr}_k\,e^{-\beta H_k}\right) \;.
\ee
 Here we used ${\rm Tr}\,\left(\prod_k e^{-\beta H_k}\right)  =
\prod_{k>0}{\rm Tr}_k\,e^{-\beta H_k}$, where $\Tr_k$ is the trace
evaluated in the Fock space of the orbital
pair $(k,\bar k)$, i.e. in the 4-dimensional space spanned by
\{$|n_k, n_{\bar
k}\rangle\}=\{|0,0\rangle,|0,1\rangle,|1,0\rangle,|1,1\rangle$\}.  In this
representation, $H_k$ is the matrix
\be\label{H-4}
\!\!\!\!\! H_k =\! \left(\matrix{G/2 & 0 & 0 & G\xi \cr
              0 & \epsilon_k-\mu & 0 & 0\cr
              0 & 0 & \epsilon_k-\mu & 0\cr
              G \xi^* &0 & 0 & 2(\epsilon_k -\mu )-G/2\cr}\right) \;.
\ee
The traces in (\ref{SPA-Z}) are easily evaluated by diagonalizing each
$H_k$ in the corresponding 4-dimensional space.  The four eigenvalues are
$\epsilon_k -\mu + \{-E_k, 0,0,E_k\}$, where
\be\label{quasi-particle}
E_k = \sqrt{(\epsilon_k-\mu-G/2)^2+G^2 |\xi|^2}
\ee
 are the familiar quasiparticle energies\footnote{One usually denotes the
self-consistent $G|\xi|$ as $\Delta$ in BCS formulation.} but now defined
for an arbitrary complex pairing field $\xi$. The trace in the subspace
($k,\bar k$) is then easily evaluated as
\begin{equation}\label{trace-k}
\begin{array}{ll}
{\rm Tr}_k e^{-\beta H_k} & =  e^{-\beta (\epsilon_k-\mu)} 4 \cosh^2 (\beta E_k/2)
\\ & = e^{-\beta (\epsilon_k-\mu)} (1+e^{-\beta E_k})
(1 + e^{\beta E_k})\;.
\end{array}
\end{equation}
The last algebraic form is convenient when dividing by $Z$ in the evaluation
of expectation values, as the reciprocal is proportional to
$f_k (1-f_k)$ where $f_k$ are the quasiparticle occupation probabilities.

An alternative way of calculating the trace is to write $H_k= (\epsilon_k
-\mu) + (\ad \; \ab){\cal H}_k \left(\a \atop \adb\right)$ where ${\cal
H}_k$ is the $2\times 2$ matrix
\be\label{H-2}
{\cal H}_k = \left(\begin{array}{cc}
        \epsilon_k-\mu-{G\over 2} & G\xi\\
        G\xi^{*} & -\epsilon_k+\mu+{G\over 2}\end{array}\right) \;,
\ee
and use the identity~\cite{fl01}
\be
{\rm Tr} \, \exp\left[(\ad \; \ab){\cal K}\left(\a \atop
\adb\right)\right] = \det \left(1 + e^{\cal K}\right)
\ee
for the matrix ${\cal K} \equiv -\beta {\cal H}_k$. The eigenvalues of
${\cal H}_k$ are just $\pm E_k$, leading again to Eq.~(\ref{trace-k}).

The complete grand-canonical partition function is given by
\be
\!\!\! Z= {\beta G\over 2\pi} \int d \xi d\xi^* e^{-\beta G |\xi|^2-\beta \sum_{k>0}
(\epsilon_k-\mu)}
\prod_{k>0} 4 \cosh^2 (\beta E_k/2) \;.
\ee
It can also be written in the form
\be
Z ={\beta G \over 2\pi}\int d\xi d\xi^{*}\;
e^{-\beta F\left[\xi,\xi^{*}\right]} \;,
\ee
where
\be\label{free-xi}
F\left[\xi,\xi^{*}\right]=G|\xi|^2+\sum_{k>0} (\epsilon_k-\mu)
-2\beta^{-1}\sum_{k>0}\ln \left[2\cosh(\beta E_k /2)\right]
\ee
is the free energy for a complex pairing field $\xi$.

The canonical partition function $Z_N={\rm Tr}_N e^{-\beta H}$ can be
calculated by a Fourier transform of the grand-canonical partition
\begin{equation}\label{canonical}
\begin{array}{ll}
\!\!\! Z_N&{\displaystyle = {1\over 2\pi i}\int_{-i\pi}^{i\pi}
d\alpha\; e^{-\alpha N} {\rm Tr}(e^{\alpha \hat{N}} e^{-\beta H})}
\vspace{.2cm}\\
&{\displaystyle ={\beta G\over 2\pi}\int d\xi d\xi^{*}\;{1\over 2\pi i}
\int_{-i\pi}^{i\pi}d\alpha\;e^{-\alpha N} e^{-\beta F[\xi, \xi^{*}]} } \;.
\end{array}
\end{equation}
The integrals in (\ref{canonical}) can be evaluated in the saddle point
approximation in both $\xi$ and $\mu$. The corresponding saddle-point
equations,  $\partial F/\partial \xi^*=0$ and $\partial F/\partial
\mu=-N$, will give the usual finite-temperature
BCS equations
\begin{equation}\label{BCS}
{1\over G}=\sum_{k>0} {\tanh\left({\beta E_k \over 2}\right)\over 2E_k} \;,
\end{equation}
and
\begin{equation}\label{BCS-N}
N=\sum_{k>0} \left(1-{\epsilon_k -\mu-{G\over 2} \over E_k} \tanh
{\beta E_k \over 2}\right) \;,
\end{equation}
where the quasi-particle energies $E_k$ are given by (\ref{quasi-particle}).
  The solutions of (\ref{BCS}) and (\ref{BCS-N}) determine the pairing gap
$\Delta=G|\xi|$ and the chemical potential $\mu$ as a function of $T$ and
particle number $N$ (the phase of $\xi$ is undetermined).

A better estimate of the canonical partition function $Z_N$ can be
obtained by a saddle-point integration in (\ref{canonical})  over
$\alpha$ for every $\xi$, but keeping the integration over $\xi$ intact.
We find
\begin{equation}\label{SPA-Z-N}
\!\!\!\! Z_N \approx 2\beta G\int_0^\infty d|\xi|\;|\xi| \left(2\pi T
{\partial^2 F\over \partial \mu^2}\right)^{-1/2} e^{-\beta
(F[\xi,\xi^{*}]+\mu N)} \;,
\end{equation}
where $\mu=\mu(N, T, \xi)$ is determined from  ${\displaystyle
{\partial F/\partial \mu}=-N}$, i.e., Eq.~(\ref{BCS-N}), and
\be
\!\!\!\! {\displaystyle \frac{\partial^2 F}{\partial \mu^2} = -\sum_{k>0}
\frac{\beta E_k (\epsilon_k-\mu-{G\over 2})^2
+\sinh(\beta E_k)G^2 |\xi|^2}{2E_k^3 \cosh^2 (\frac{\beta E_k}{2})}}\;.
\ee
In Eq.~(\ref{SPA-Z-N}) we have carried out explicitly the integral over
the phase of the pairing field $\xi$ since the integrand was only a
function of $|\xi|$.

\subsubsection{Moment of inertia}\label{sec:inertia-xi}

Our model (\ref{Hamiltonian}) describes a non-rotationally invariant
Hamiltonian, and we can use the formalism of Section \ref{sec:inertia} to
estimate the moment of inertia $I$ in terms of the intrinsic moments
$I_{ii}$ (see Eq.~(\ref{inertia3})). Rather then using
Eqs.~(\ref{inertia}) and (\ref{inertia-tensor}) with the full Hamiltonian $H$,
we first apply an SPA representation similar to (\ref{SPA}) but for the
cranked Hamiltonian in (\ref{free}), and then calculate the intrinsic
moments from $I_{ii}= \partial^2 F(\beta,\vec\omega)/\partial
\omega_i^2|_{\omega=0}$. If this is done starting from the canonical
partition function in (\ref{free}), we obtain the following expression
\begin{equation}\label{inertia-int}
 I_{ij}={\int d\xi d\xi^{*}\;
\int_{-i\pi}^{i\pi}d\alpha\;e^{-\alpha N} e^{-\beta F[\xi, \xi^{*}]}{\cal
I}_{ij} (\xi) \over
\int d\xi d\xi^{*}\;{1\over 2\pi i}
\int_{-i\pi}^{i\pi}d\alpha\;e^{-\alpha N} e^{-\beta F[\xi, \xi^{*}]}}\;,
\end{equation}
where $F[\xi,\xi^*]$ is the free energy (\ref{free-xi}) and
\be\label{inertia-xi}
{\cal I}_{ij}(\xi) = \int_0^\beta d\tau \,
{{\rm Tr} [e^{-\beta H_\xi} J_i(\tau)J_j(0)] /\, {\rm Tr}\,
[e^{-\beta H_\xi}]}\;,
\ee
 where $J_i(\tau) = e^{\tau H_\xi} J_i e^{-\tau H_\xi}$ and $H_\xi\equiv
\sum_{k>0} H_k$ is the mean-field Hamiltonian in a pairing field $\xi$.
 Expression (\ref{inertia-xi}) is analogous to (\ref{inertia}) and
 (\ref{inertia-tensor}), except that the Hamiltonian $H'$ in those
expressions is now replaced by $H_\xi$.

   The integrals over $\alpha$ in (\ref{inertia-int}) can be done in the
saddle-point as before to obtain the final expression for the
intrinsic moments
\begin{equation}\label{inertia-xi-N}
  I_{ij}={\int_0^\infty d\vert\xi\vert\;\vert\xi\vert
(\partial^2 F/\partial \mu^2)^{-1/2} e^{-\beta (F[\xi,\xi^\ast]+\mu N)}
{\cal I}_{ij} (\xi) \over
\int_0^\infty d\vert\xi\vert\;\vert\xi\vert
(\partial^2 F/\partial \mu^2)^{-1/2} e^{-\beta (F[\xi,\xi^\ast]+\mu N)}}\;.
\end{equation}

 It remains to calculate the moments ${\cal I}_{ij}(\xi)$.  The operator
$J_z$ leaves the $(k,\bar k)$ subspace invariant, but
the operators $J_x$ and $J_y$ connect different subspaces $k$ and
 $k'$, so the trace in (\ref{inertia-xi}) is to be evaluated in a
16-dimensional space. This is most conveniently done in the quasiparticle
representation. The transformation from deformed single-particle states to
the quasi-particle states is achieved through a Bogoliubov transformation
\begin{equation}\label{Bogoliubov}
\left(\begin{array}{c}\alpha_k\\ \alpha_{\bar{k}}^\dagger\end{array}\right)
=U_k\left(\begin{array}{c}a_k\\a_{\bar{k}}^\dagger\end{array}\right)=
\left(\begin{array}{cc}u_k & -v_k\\ v_k^{*} & u_k\end{array}
\right)\left(\begin{array}{c}a_k\\a_{\bar{k}}^\dagger\end{array}\right) \;,
\end{equation}
 where $u_k$ is real and $v_k$ is complex, and $u_k^2+|v_k|^2=1$ to
preserve the fermionic commutation relations. Relations
(\ref{Bogoliubov}) imply
\be\label{bar-k}
u_{\bar k} = u_k \;;\;\;\; v_{\bar k} = - v_k\;\;\;{\rm for}\;\;k>0 \;.
\ee
 The parameters $u_k,v_k$ are chosen such that $H_k$ in
Eq.~(\ref{pairing-k}) is diagonal in the quasi-particle representation, i.e.,
\be
U^\dagger_k {\cal H}_k U_k = \left(\begin{array}{cc} E_k & 0\\ 0 &
-E_k\end{array} \right)\;,
\ee
where ${\cal H}_k$ is the $2\times 2$ matrix (\ref{H-2}) and $E_k$ are
the eigenvalues (\ref{quasi-particle}) of ${\cal H}_k$.  The solution is
\begin{eqnarray}\label{u-v}
u_k^2 &= &{1\over 2}\left(1+ {\epsilon_k-\mu-G/2 \over E_k}\right)\\ \nonumber
|v_k|^2&=& {1\over 2}\left(1 - {\epsilon_k-\mu-G/2 \over E_k}\right) \;,
\end{eqnarray}
and $\arg v_k =  \arg \xi$.
The Hamiltonian $H_k$ is now given by $H_k= (\epsilon_k-\mu)
+E_k(\alpha^\dagger_k \alpha_k - \alpha_{\bar k} \alpha^\dagger_{\bar k})$, and
\be\label{H-qp}
H_\xi = \sum_{k>0} E_k (\alpha^\dagger_k \alpha_k + \alpha^\dagger_{\bar
k} \alpha_{\bar k}) +\sum_{k>0} (\epsilon_k-\mu - E_k) \;.
\ee

 Expressing $J_i$ in the quasi-particle representation, and using
(\ref{H-qp}), we can calculate the intrinsic moments in closed form (see
Appendix A). The final result is
\begin{widetext}
\begin{equation}\label{inertia-qp}
\begin{array}{ll}{\displaystyle
\!\!\!\! \! {\cal I}_{ij}(\xi) =}&{\displaystyle \sum_{k, l>0}
\left[ (\langle k\vert j_i\vert l\rangle
\langle k\vert j_j\vert l\rangle^{*} + \langle k\vert j_i\vert -l\rangle
\langle k\vert j_j\vert -l\rangle^{*}) + c.c. \right]}
 \left\{(u_k u_l +v_k v_l)^2{f_l-f_k \over E_k - E_l} +
(u_k v_l-v_k u_l)^2{1-f_k-f_l \over E_k +E_l}\right\} \;,
\end{array}
\end{equation}
where
\be
{\displaystyle f_k={1\over 1+e^{\beta E_k}}}
\ee
are the quasi-particle occupations. $u_k$ and $v_k$ are still given by
(\ref{u-v}), except that now we have chosen
\be
u_k, v_k>0\;\;\;{\rm for}\;\;k>0 \;,
\ee
and $u_{\bar k}, v_{\bar k}$ are still given by (\ref{bar-k}).
 Eq.~(\ref{inertia-qp}) is the finite-temperature generalization
  of the Belyaev formula~\cite{be61}.

Eq.~(\ref{inertia-qp}) can be rewritten by separating out the
contribution from the $k=l$ terms in the sum.  Using $(f_l-f_k)/(E_k -
E_l)\stackrel{k=l}{\longrightarrow}
 -{\partial f_k/ \partial E_k}$, we obtain (for $i=j$)
\begin{equation}\begin{array}{ll}
{\cal I}_{ii}(\xi)=&{\displaystyle 2\sum_{k>0}(|\langle k | j_i |
k\rangle |^2+
|\langle k | j_i | -k\rangle |^2)\left(-{\partial f_k \over \partial E_k}
\right)+
2\sum\limits_{k, l>0} \left(\vert \langle k\vert j_i\vert l
\rangle\vert^2 + \vert \langle k\vert j_i\vert -l\rangle\vert^2\right)}\\&
{\displaystyle \times \left\{(u_k u_l +v_k v_l)^2{f_l-f_k \over E_k - E_l} +
(u_k v_l-v_k u_l)^2{1-f_k-f_l \over E_k +E_l}\right\}}\;.
\end{array}
\end{equation}
In particular, $\langle k\vert j_z \vert l\rangle=0$ for $k\ne l$, and
$\langle k\vert j_x \vert k\rangle=0$. Therefore,
 the moment of inertia around an axis parallel to the symmetry axis $z$
(non-collective rotation) is given by
\begin{equation}
{\cal I}_{zz}(\xi)=2\sum_{k>0} \vert\langle k\vert j_z \vert k\rangle\vert^2
\left(-{\partial f_k \over \partial E_k}\right) \;,
\end{equation}
while the moment of inertia around an axis $x$ perpendicular to the
symmetry axis (collective rotation) is given by
\begin{equation}\label{I-xx}
\begin{array}{ll}
{\cal I}_{xx}(\xi)=&{\displaystyle 2\sum_{k>0} \vert\langle k\vert j_x \vert
-k\rangle\vert^2 \left(-{\partial f_k \over \partial E_k}\right)+
2\sum\limits_{k, l>0} \left(\vert \langle k\vert j_x\vert l
\rangle\vert^2 + \vert \langle k\vert j_x\vert -l\rangle\vert^2\right)}\\&
{\displaystyle \times \left\{(u_k u_l +v_k v_l)^2{f_l-f_k \over E_k - E_l} +
(u_k v_l-v_k u_l)^2{1-f_k-f_l \over E_k +E_l}\right\}} \;.
\end{array}
\end{equation}

 In the limit $T\rightarrow 0$ (but $\Delta>0$), $-\partial f_k/\partial
E_k \to \delta(E_k)$, and since $E_k>0$, $I_{zz}(\xi)=0$. Also $f_k \to
0$ and Eq.~(\ref{I-xx}) reduces to
\begin{equation}\label{Belyaev}
\begin{array}{ll}
 I_{xx}(\xi)&{\displaystyle = 2\sum_{k\ne l>0}
\left(\vert \langle k\vert j_x\vert l
\rangle\vert^2 + \vert \langle k\vert j_x\vert -l\rangle\vert^2\right)
{(u_k v_l-v_k u_l)^2 \over E_k +E_l}}
{\displaystyle =\sum_{k, l} \vert \langle k\vert j_x\vert l\rangle\vert^2
{(u_k v_l-v_k u_l)^2 \over E_k +E_l} } \;.
\end{array}
\end{equation}
\end{widetext}
Eq.~(\ref{Belyaev}) is known as the Belyaev formula~\cite{be61}; it
produces a moment of inertia that is suppressed relative to the
rigid-body value.

\section{Number-Parity Projection}\label{sec:number-parity}

  In the relations we derived in the previous Section for the partition
function and moment of inertia, the number of particles is fixed only on
average, and odd-even effects cannot be reproduced. Here we go through
the same derivation steps but include now the number-parity projection
operator $P_\eta$.  The resulting formulas will exhibit explicit
terms depending on the number parity.

\subsection{Partition function}

  The projected partition function (\ref{projected-partition}) introduces
the operator $e^{i\pi \hat N}$. In the SPA
\begin{eqnarray}
 {\rm Tr}\, \left( e^{i\pi \hat N} e^{-\beta H'}\right) &\!\! \!= \!\!\!&
{\displaystyle {\beta G\over 2\pi} \int} d \xi d\xi^*
e^{-\beta G |\xi|^2} \nonumber
\\ \times \prod_{k>0}&{\rm Tr}_k& \left[ e^{i\pi (\ad\a + \adb\ab)} e^{-\beta
H_k}\right]  \;.
\end{eqnarray}

 Within each subspace $(k,\bar k)$, the operator $e^{i \pi N}$ changes the
 sign of the two vectors $|0,1\rangle,|1,0\rangle$, but leaves the sign of
 $|0,0\rangle,|1,1\rangle$ unchanged. The matrix representing  $e^{i\pi
(\ad\a + \adb\ab)}$ is then
\be
     \left(\matrix{1 & 0 & 0 & 0 \cr
              0 & -1  & 0 & 0\cr
              0 & 0 & -1 & 0\cr
          0 &0 & 0 &1\cr}\right) \;.
\ee
 The transformation that diagonalizes $H_k$ in Eq.~(\ref{H-4}) leaves this
matrix invariant, and therefore the trace is now given by
\begin{eqnarray}\label{trace-k-N}
&{\rm Tr}_k\,\left[ e^{i\pi (\ad\a + \adb\ab)} e^{-\beta H_k}\right]
= e^{-\beta (\epsilon_k-\mu)} 4 \sinh^2 (\beta E_k/2) \nonumber
\\ & =  e^{-\beta (\epsilon_k-\mu)} (1-e^{-\beta E_k})
(e^{\beta E_k}-1).
\end{eqnarray}
 The projected grand-canonical partition function  $Z_\eta \equiv {\rm
 Tr}\, \left(P_\eta e^{-\beta H'}\right)$ is now calculated from
(\ref{trace-k}) and (\ref{trace-k-N}) to be
\be\label{projected-Z}
Z_\eta = {\beta G\over 4\pi} \int  d \xi d\xi^* e^{-\beta F[\xi,\xi^*]}
 \left(1 + \eta \prod_{k>0} \tanh^2{\beta E_k \over 2}\right) \;.
\ee
Notice that the integrand in (\ref{projected-Z}) has the form
 of Eq.~(\ref{projected-partition}) when applied to the Hamiltonian
$H_\xi = \sum_{k>0} H_k$ at a fixed pairing field $\xi$. Indeed
\be\label{pi-N}
\langle e^{i\pi \hat N}\rangle_\xi \equiv {{\rm Tr}\,\left(e^{i\pi \hat
N} e^{-\beta H_\xi}\right) \over {\rm Tr}\,e^{-\beta H_\xi}} =
\prod_{k>0} \tanh^2{\beta E_k \over 2}\;.
\ee
This projected partition can also be written as
$Z_\eta= {\beta G \over 2\pi}
\int d\xi d\xi^{*}\;e^{-\beta F_\eta\left[\xi,\xi^{*}\right]}$
where
\be
\!\!\!\!\!\!F_\eta\left[\xi,\xi^{*}\right]=F\left[\xi,\xi^{*}\right]-\beta^{-1}
\ln \left({1+\eta\prod_{k>0}\tanh^2{\beta E_k \over 2} \over 2} \right)
\ee
is the number-parity projected free energy. Proceeding as in Section
\ref{sec:partition},
we can derive (in the saddle point approximation)
number-parity projected BCS equations
\begin{equation}\label{NP-BCS}
\begin{array}{l}{\displaystyle
{1\over G}=\sum_{k>0} \left({\tanh\left({\beta E_k \over 2}\right)\over 2E_k}
+{C_\eta \over E_k \sinh(\beta E_k)}\right)}
\end{array} \;,
\end{equation}
and
\begin{equation}
\begin{array}{l}{\displaystyle
\!\!\!\!\! N=\sum_{k>0} \left[1-{\epsilon_k -\mu-{G\over 2} \over E_k} \left(\tanh
{\beta E_k \over 2}+{2C_\eta \over \sinh(\beta E_k)}\right)\right]}
\end{array} \;,
\end{equation}
where
\be
{\displaystyle C_\eta={\eta \prod_{k>0} \tanh^2\left({\beta E_k
\over 2}\right) \over 1+\eta \prod_{k>0} \tanh^2\left({\beta E_k\over 2}
\right)}}\;.
\ee

It is interesting to take the
$T\rightarrow 0$ limit for the above equations.
For $\eta=+1$, they simply become the usual $T=0$ BCS equations. However,
for the odd projection
$\eta=-1$, we find (assuming there are no degeneracies and $\Delta=G|\xi|>0$)
\begin{equation}\label{BCS-odd}
\begin{array}{l}{\displaystyle
{1\over G}=\sum_{k>0} {1\over 2E_k} -{1\over 2E_{k_0}}}\\{\displaystyle
N=\sum_{k\ne k_0 >0} 2|v_k|^2 +1}\;,
\end{array}
\end{equation}
where $|v_k|^2$ is given by (\ref{u-v}) and
$E_{k_0}$ is the lowest quasi-particle energy (corresponding to
$\epsilon_{k_0}$
closest to $\mu$). In deriving (\ref{BCS-odd}), we have used the limit
$C_{-1}\stackrel{\beta\rightarrow\infty}{\longrightarrow}
-e^{\beta E_{k_0}}/4$. Eqs.~(\ref{BCS-odd}) reproduce what is known as the
blocking effect, since one level $k=k_0$ is ``blocked'' and does not
contribute to the sum over $k$.

\begin{figure}[bth]
\centerline{\epsfig{figure=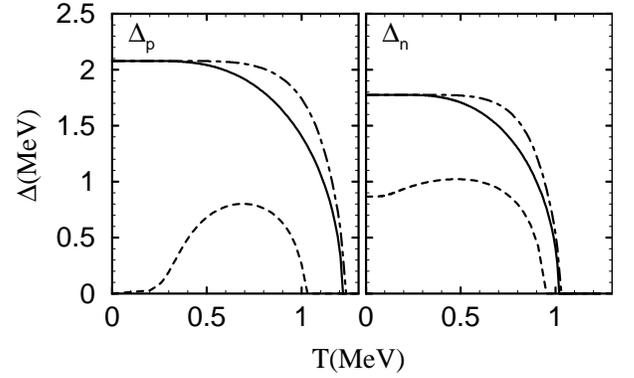,width=8 cm,clip=}}
\vspace{3 mm}
\caption{\label{fig1}
 The pairing gap $\Delta$ versus temperature $T$ for protons (left) and
neutrons (right). The solid lines are the solution of the BCS equations
(\protect\ref{BCS}) for $^{56}$Fe. The dotted-dashed lines and the dashed
lines are respectively the solution to the number-parity projected BCS
equations (\protect\ref{NP-BCS}) for $\eta=1$ and $\eta=-1$.
 The deformation parameter is taken to be $\beta_2=0.24$ and the pairing
strengths are determined from the zero-temperature BCS to reproduce the
experimental values of the gaps (using the second difference formula of
Ref.~\protect\cite{bm69}).
}
\end{figure}

 In Fig.~\ref{fig1} we display the solution of the number-parity
projected BCS equations for a Hamiltonian corresponding to the
nucleus $^{56}$Fe.  The pairing gap $\Delta$ is shown as the solid line
as a function of temperature $T$.  For the particle-number projected
BCS equations (\ref{NP-BCS}), the solutions for even
and odd particle numbers are shown
as the dot-dashed and dashed lines, respectively.
The proton gap $\Delta_p$ and the neutron gap $\Delta_n$ are shown in
the left and right panels.
Note the strong suppression of the gap for the odd projection. Our
results for the projected gap are qualitatively similar to
Ref.~\cite{ba99,fl01}, but the formulas are quite different.

The number-parity projected partition function at a fixed average number
of particles is given by an equation similar to (\ref{SPA-Z-N})
except that $F$ is replaced by $F_\eta$.

\subsection{Moment of inertia}\label{sec:NP-inertia}

   The number-parity projected moment of inertia can be calculated as in
 Section \ref{sec:inertia-xi}, but now starting from the number-parity
projected free energy in the presence of a rotational field $\vec\omega$.
The result is
\begin{equation}\label{NP-inertia}
 I_{ij}^\eta={\int_0^\infty d\vert\xi\vert\;\vert\xi\vert
(\partial^2 F_\eta/\partial \mu^2)^{-1/2} e^{-\beta (F_\eta[\xi,\xi^*] +\mu N)}
{\cal I}_{ij}^\eta (\xi) \over
\int_0^\infty d\vert\xi\vert\;\vert\xi\vert
 (\partial^2 F_\eta/\partial \mu^2)^{-1/2} e^{-\beta (F_\eta[\xi,\xi^*]
+\mu N)}} \;,
\end{equation}
where
\be
 I_{ij}^\eta(\xi)= \int_0^\beta d\tau\;{{\rm Tr}\, \left[P_\eta e^{-\beta
H_\xi} J_i(\tau)J_j(0)\right] \over {\rm Tr}\, \left[P_\eta
e^{-\beta H_\xi} \right]} \;.
\ee
 Thus we need to calculate the projected value $\langle
J_i(\tau)J_j(0)\rangle_\eta$.
The odd-even number projection can be carried out for any operator
 ${\cal O}$ using (\ref{projected-observable}) at a fixed pairing field
$\xi$ together with (\ref{pi-N})
\be
 \langle {\cal O}\rangle_\eta =  {\langle {\cal O}\rangle + \eta \left(
\prod_{k>0}
\tanh^2{\beta E_k \over 2}\right) \langle {\cal O}\rangle_\pi \over 1+ \eta
\left( \prod_{k>0} \tanh^2{\beta E_k \over 2}\right)}.
\ee
 In general, $ \langle {\cal O}\rangle_\pi = {{\rm Tr}\, \left({\cal O}
 e^{i \pi\hat N} e^{-\beta H_\xi}\right)/ {\rm Tr}\, \left( e^{i \pi\hat
N} e^{-\beta H_\xi}\right)}$ will have the
same form as $\langle {\cal O}\rangle$ but with $e^{\pm \beta E_k}$
replaced by $-e^{\pm \beta E_k}$.

\begin{widetext}
For the intrinsic moment of inertia we find
\begin{equation}\label{NP-inertia-xi}
{\cal I}_{ij}^\eta(\xi)={\int_0^\beta d\tau\;\langle J_i(\tau)J_j(0)\rangle
+ \eta \prod_{k>0}\tanh^2 {\beta E_k \over 2}
\int_0^\beta d\tau\;\langle J_i(\tau)J_j(0)\rangle_\pi
\over 1+\eta \prod_{k>0} \tanh^2 {\beta E_k \over 2}} \;,
\end{equation}
where $\int_0^\beta d\tau\,\langle J_i(\tau)J_j(0)\rangle$ is given by (\ref{inertia-qp}) and
$\int_0^\beta d\tau\,\langle J_i(\tau)J_j(0)\rangle_\pi$ is obtained
from the expression
for $\int_0^\beta d\tau\,\langle J_i(\tau)J_j(0)\rangle$ by the substitution
\be
f_k \rightarrow \tilde{f_k}={1\over 1-e^{\beta E_k}}\;.
\ee

We now inspect the $T \to 0$ limit of Eq.~(\ref{NP-inertia-xi}). For the
even number-parity projected inertia ${\cal I}^{\eta=1}_{zz} \to 0$ and
${\cal I}^{\eta=1}_{xx}$ is the same as in Eq.~(\ref{Belyaev}).
For the odd number-parity projected moment of inertia, we have in the
limit $T\to 0$
\begin{equation}
{\cal I}^{\eta=-1}_{zz}\to \beta |\mu_0 |^2 \;,
\end{equation}
where $|k_0\rangle=|q,\;\mu_0\rangle$ ($E_{k_0}$ is the lowest
quasi-particle energy),  and
\begin{equation}
\begin{array}{ll}
{\cal I}^{\eta=-1}_{xx}\to &{\displaystyle 2\sum\limits_{k\ne k_0 >0}
(|<k|j_x |k_0 >|^2+|<k|j_x | -k_0 >|^2)
\left\{\frac{(u_k u_{k_0}+v_k v_{k_0})^2}{E_k -E_{k_0}}+
\frac{(u_k v_{k_0}-v_k u_{k_0})^2}{E_k +E_{k_0}}\right\}}
\vspace{.0cm}\\
&{\displaystyle +2\sum_{k,l\ne k_0 >0}(|<k|j_x |l>|^2+|<k|j_x |-l>|^2)
{(u_k v_l-v_k u_l)^2\over E_k +E_l}+\beta |<k_0 |j_x |-k_0 >|^2}\;.
\end{array}
\end{equation}
\end{widetext}

The number-parity projected moments of inertia $I_{ij}^\eta$ are computed by
a numerical integration of Eq.~(\ref{NP-inertia}), using Eq.~(\ref{inertia-qp})
and (\ref{NP-inertia-xi}).  For the final result, the contributions from the
three principal axes must be combined according to Eq.~(\ref{inertia3}).

\section{Comparison to SMMC results}\label{sec:results}

 We have used our formulas to study the moment of inertia $I$ versus
temperature for nuclei in the iron region and compared the results with
SMMC calculations. The effective Hamiltonian is defined in the full
 $pfg_{9/2}$ shell with single-particle
energies determined by a Woods-Saxon potential plus spin-orbit term. The
interaction includes $T=1$ monopole pairing and multipole-multipole
interaction terms (up to hexadecupole)~\cite{na97}.   In SMMC we
calculate $\langle J_z^2\rangle$ as an observable and then find $I$  from
\be\label{SMMC-I}
{I\over \hbar^2} = \beta \langle J_z^2\rangle \;.
\ee

\begin{figure}[bth]
\centerline{\epsfig{figure=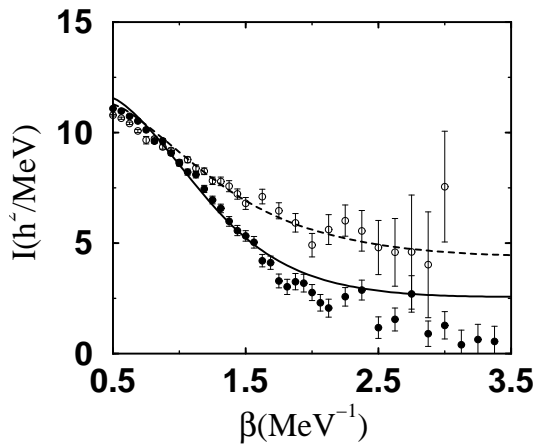,width=7 cm,clip=}}
\vspace{3 mm}
\caption{\label{fig2}
 Moment of inertia (characterizing the spin-cutoff distribution
(\ref{spin-cutoff})) versus inverse temperature $\beta$ for an even-even
nucleus ($^{56}$Fe) and an odd-even nucleus ($^{55}$Fe).  The SMMC
results calculated from (\protect\ref{SMMC-I}) are shown by the solid
circles for $^{56}$Fe and the open circles for $^{55}$Fe. The results of
the model discussed in this work (which includes fluctuations of the
pairing field and number-parity projection) are shown by solid line for
$^{56}$Fe and dashed line for $^{55}$Fe. For protons we use even
number-parity projection, while for neutrons we use even (odd)
number-parity projection for $^{56}$Fe ($^{55}$Fe).  In the model we use
a deformation of $\beta_2=0.14$ and pairing strengths of $G_p=0.42$ and
$G_n=0.36$.}
\end{figure}

 Significant odd-even effects are observed in the SMMC
 calculations~\cite{spin-proj}, see, e.g., in Fig.~\ref{fig2}
 where the SMMC thermal
moment of inertia of the even-even nucleus $^{56}$Fe (solid circles) is
compared with the SMMC moment of inertia of the odd-even nucleus
$^{55}$Fe (open circles). We see stronger suppression at the lower
temperatures in the even-even case.

The parameters of the model we discuss in the present work are the
deformation $\beta_2$ and the pairing strengths $G_p$, $G_n$ for protons
and neutrons, respectively. We have chosen $\beta_2=0.14$ and computed
$G_p$ and $G_n$ from the experimental values of $\Delta_p$ and $\Delta_n$
(using zero-temperature BCS). In our model, we include fluctuations in
the pairing gap and we have found it necessary to reduce the BCS values
of $G$ by about 20\% to reproduce correctly the high temperature behavior
of $I$. We then fixed $\beta_2=0.14$, $G_p=0.42$ and $G_n=0.36$
through the whole mass region, and calculated the number-parity projected
moment of inertia $I^{\eta}$ (with fluctuations in the pairing order
parameter included) using Eqs. (\ref{inertia3}), (\ref{NP-inertia}),
(\ref{NP-inertia-xi}) and (\ref{inertia-qp}). Fig.~\ref{fig2} shows the
results for $I^\eta$ in $^{56}$Fe (solid line) and in $^{55}$Fe (dashed
line). We find that the inclusion of a number parity projection allows
us describe reasonably well the odd-even effect observed in
the SMMC moment of inertia. We note that an exact particle-number
projection (for both protons and neutrons) is used in the SMMC method.

\begin{figure}[bth]
\centerline{\epsfig{figure=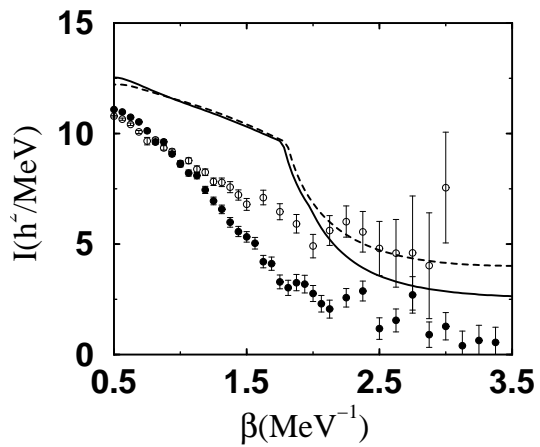,width=7. cm,clip=}}
\vspace{3 mm}
\caption{\label{fig3}
Similar to Fig.~\protect\ref{fig2} but the lines correspond to the
number-parity projected moments of inertia calculated at the solutions
$\xi$ of the corresponding number-parity BCS equations
(\protect\ref{NP-BCS}). The deformation and pairing parameters are the
same as in Fig.~\protect\ref{fig2}.
}
\end{figure}

  To demonstrate the importance of fluctuations in the pairing order
parameter, we show in Fig.~\ref{fig3} the number-parity projected moments
of inertia $I^\eta(\xi)$ evaluated at the corresponding solutions $\xi$
to the number-parity projected BCS equations (\ref{NP-BCS}) but without
the inclusion of fluctuations. $I^\eta(\xi)$ is calculated from
(\ref{NP-inertia-xi}) and averaging over principal axes.  We use the same
deformation and pairing parameters as in the theory with fluctuations.
These results demonstrate that we cannot obtain good agreement with the
microscopic SMMC calculations without including fluctuations.

\begin{figure}[bth]
\centerline{\epsfig{figure=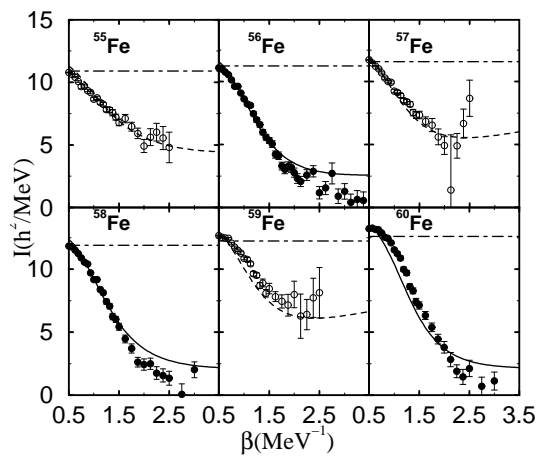,width=7. cm,clip=}}
\vspace{3 mm}
\caption{\label{fig4}
Systematics of the moment of inertia versus $\beta$ for a series of iron
isotopes. The symbols (solid circles for even-mass isotopes and open
circles for odd-mass isotopes) are the SMMC results calculated from
(\protect\ref{SMMC-I}). The lines describe the results of our model
(including fluctuations in the pairing field). The solid lines are the
even number-parity projection while the dashed lines describe the odd
number-parity projection. The dotted-dashed lines are the rigid-body
moment of inertia.
We use $\beta_2=0.14$ and  $G_p=0.42$, $G_n=0.36$.
}
\end{figure}

 In Fig.~\ref{fig4} we present a systematic study of the moment of inertia
for even and odd iron isotopes from $^{55}$Fe to $^{60}$Fe. We show the
moment of inertia versus $\beta$, comparing the SMMC results (symbols
with statistical error bars) to our model results (solid lines for even
isotopes and dashed lines for odd isotopes).  The model includes
fluctuations in the pairing fields, and we have used a fixed set of
deformation and pairing parameters. Despite its simplicity, the results
of our model agree well with the full microscopic SMMC calculation.

\begin{figure}[t!]
\centerline{\epsfig{figure=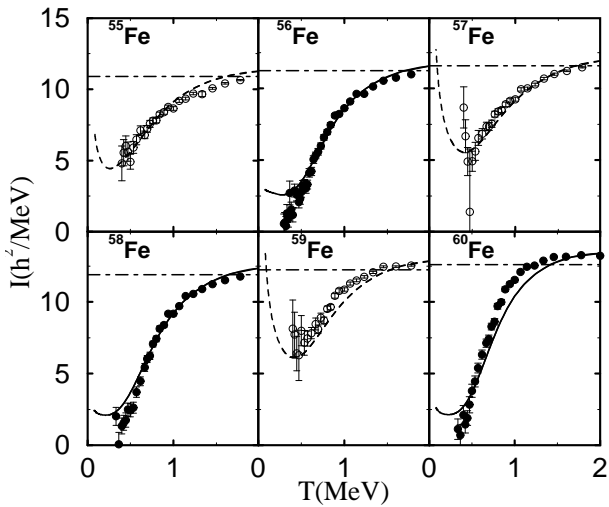,width=8 cm,clip=}}
\caption{\label{fig5}
 As in Fig.~\protect\ref{fig4}, except that the moments of inertia are
shown  versus temperature.
}
\end{figure}

\begin{figure}[bth]
\vspace{5 mm}
\centerline{\epsfig{figure=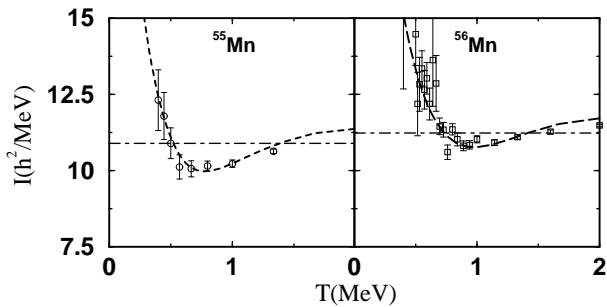,width=8 cm,clip=}}
\caption{\label{fig6}
 The moments of inertia versus temperature for the odd-even nucleus
$^{55}$Mn and
 the odd-odd nucleus $^{56}$Mn. The notation as in Fig.~\protect\ref{fig4}.
}
\end{figure}

 In Fig.~\ref{fig5} we show the same results as in Fig.~\ref{fig4} but
plotted as a function of temperature.  In general, the thermal moment of
inertia is seen to decrease with decreasing temperature with stronger
suppression in the even isotopes. Furthermore, for the odd iron isotopes
(in which an odd number-parity projection is used for neutrons), the
moment of inertia starts to rise at low temperatures. Indeed, according to (\ref{SMMC-I})
\be
I \to {1\over 3} \beta J (J+1) \;,
\ee
 for large $\beta $ where $J$ is the ground-state spin. In an odd-even
nucleus, $J\neq 0$ and $I$ increases linearly with $\beta$ at large
$\beta$. In the model this rise at low temperatures can be understood
to be the effect of the unpaired neutron. Using the projected moment
of inertia for odd number of particles, we find in the limit of
 large $\beta$
\be\label{odd-0}
I^{\eta=-1} \to {1\over 3}\beta( \mu_0^2 + |\langle k_0 | j_x|-k_0\rangle|^2)\;,
\ee
where the second term in (\ref{odd-0})
contributes only for $\mu_0=1/2$. It is interesting to
note that a similar odd-even effect was found in the spin susceptibility
of ultra-small metallic particles (nanoparticles)~\cite{dl00}.  The rise
of the spin susceptibility at low temperatures for an odd number of
electrons is known as re-entrant behavior and survives in the
fluctuation-dominated regime.

 Finally, in Fig.~\ref{fig6} we show the moment of inertia versus
temperature for an odd-even nucleus $^{55}$Mn (left panel) and for an
odd-odd nucleus
 $^{56}$Mn (right panel).
 We have used the same deformation and pairing strength parameters as for
the iron nuclei. Again we find good agreement between
the results of the model (dashed lines) and the microscopic SMMC
calculations (symbols).

\section{Conclusion}

   We have presented a simple model to calculate the nuclear moment of
inertia at finite temperature. The model includes quadrupolar deformation
of the single-particle field and pairing interaction.  The pairing
interaction is treated beyond the mean-field BCS limit by including
static fluctuations in the pairing order parameter at finite temperature.

    In the fluctuation-dominated regime, finite temperature signatures of
 pairing correlations are often observed as odd-even effects (in particle
number)~\cite{nano01}. Such signatures are usually not seen in the
grand-canonical ensemble, but only after particle-number projection is
implemented (such as in SMMC). In this work we have used number-parity
projection instead of exact particle-number projection and showed that
the odd-even effects in the moment of inertia seen in the microscopic
SMMC calculations can be well reproduced. The advantage of such
number-parity projection is its simplicity, which allows us to express
the projected moment of inertia in essentially closed form (except for an
integral over the pairing field).

  The simple model developed here is useful in estimating the spin
distribution of nuclear level densities at not-too-low temperatures for
which the spin-cutoff model usually works well.  The spin-cutoff model
depends on a single parameter -- the moment of inertia, and thus
our model is useful
for global estimates of the spin distribution below the neutron
separation energy where the moment of inertia may deviate from its
rigid-body value.

\section*{Acknowledgment}

  We thank R. Vandenbosch for discussions. 
 Y.A. would like to acknowledge the hospitality of the Institute of
Nuclear Theory in Seattle where part of this work was completed.  This
work is supported by the DOE under Grants DE-FG-02-91ER40608 and
DE-FGO3-00ER41132.

\begin{widetext}

\section*{Appendix A: finite-temperature generalization of the
Belyaev formula for the moment of inertia tensor}

 Here we derive Eq.~(\ref{inertia-qp}).  Since $H_\xi$ is diagonal in the
quasi-particle representation (see Eq.~(\ref{H-qp})), we have
\be\label{qp-tau}
\begin{array}{l}\alpha_k^\dagger(\tau)\equiv e^{\tau H_\xi} \alpha_k^\dagger
e^{-\tau H_\xi}=e^{\tau E_k} \alpha_k^\dagger\\
\alpha_k(\tau)\equiv e^{\tau H_\xi} \alpha_k e^{-\tau H_\xi}=e^{-\tau
E_k} \alpha_k \;.
\end{array}
\ee

Using (\ref{qp-tau}) and the Bogoliubov transformation
(\ref{Bogoliubov}), we obtain
\begin{equation}
  \begin{array}{rcl}
    J_i(\tau) & = & \sum\limits_{k, l} \langle k \vert j_i \vert l
    \rangle a^\dagger_k(\tau) a_l(\tau)  =  \sum\limits_{k, l}
    \langle k \vert j_i \vert l \rangle (u_k \alpha^\dagger_k(\tau) +
    v_k\alpha_{\tilde k}(\tau)) (u_l \alpha_l(\tau) + v_l
    \alpha^\dagger_{\tilde l}(\tau)) \\ & = & \sum\limits_{k, l} \langle
    k \vert j_i \vert l \rangle \left [
        e^{\tau(E_k - E_l)} u_k u_l\alpha^\dagger_k\alpha_l
        + e^{\tau(E_{\tilde l} - E_{\tilde k})} v_k v_l \alpha_{\tilde k}
        \alpha^\dagger_ {\tilde l}
        + e^{\tau(E_k + E_{\tilde l})} u_k v_l\alpha^\dagger_k
        \alpha^\dagger_{\tilde l}
        + e^{-\tau(E_{\tilde k} + E_l)} v_k u_l \alpha_{\tilde k}
        \alpha_l
    \right ].
  \end{array}
\end{equation}
The spin response function is then given by
\begin{eqnarray}
    \langle J_i(\tau)J_j(0)\rangle = & \sum\limits_{k,l,m,n} \langle k
    | j_i | l \rangle \langle m | j_j | n \rangle
     \times \left [
        e^{\tau(E_k - E_l)} u_k u_l u_m u_n \langle \alpha^\dagger_k\alpha_l
        \alpha^\dagger_m\alpha_n\rangle + e^{-\tau(E_k - E_l)} v_k v_l u_m u_n
        \langle \alpha_{\bar{k}}\alpha_{\bar{l}}^\dagger
        \alpha^\dagger_m\alpha_n\rangle \right. \nonumber \\
    & +
        e^{\tau(E_k - E_l)} u_k u_l v_m v_n \langle \alpha^\dagger_k\alpha_l
        \alpha_{\bar{m}}\alpha_{\bar{n}}^\dagger\rangle + e^{-\tau(E_k - E_l)}
        v_k v_l v_m v_n \langle \alpha_{\bar{k}}\alpha_{\bar{l}}^\dagger
        \alpha_{\bar{m}}\alpha_{\bar{n}}^\dagger\rangle \nonumber \\
    & + \left .
        e^{\tau(E_k+E_l)}u_k v_l v_m u_n \langle \alpha^\dagger_k
        \alpha_{\bar{l}}^\dagger\alpha_{\bar{m}}\alpha_n\rangle +
        e^{-\tau(E_k+E_l)}v_k u_l u_m v_n \langle\alpha_{\bar{k}}\alpha_l
        \alpha_m^\dagger\alpha_{\bar{n}}^\dagger\rangle
      \right ]\;.
\end{eqnarray}
 Using Wick's theorem in the quasi-particle representation and
$f_k=f_{\bar k}$, we find
\begin{eqnarray}\label{j-j-2}
    \langle J_i(\tau)J_j(0)\rangle & = & \sum_k\{\langle k | j_i | k\rangle
    [u_k^2 f_k+v_k^2(1-f_k)]\}\times\sum_m\{\langle m | j_j | m \rangle
    [u_m^2 f_m+v_m^2(1-f_m)]\} \nonumber\\
    && + \sum\limits_{k,l}\langle k | j_i | l\rangle\langle l | j_j | k\rangle
    \times \left[e^{\tau(E_k-E_l)}u_k^2 u_l^2f_k(1-f_l)+e^{-\tau(E_k-E_l)}
        v_k^2 v_l^2 (1-f_k)f_l\right . \nonumber\\
    & & \hspace{3 cm} + \left .
        e^{\tau(E_k+E_l)}u_k^2 v_l^2 f_k f_l+e^{-\tau(E_k+E_l)}v_k^2 u_l^2
        (1-f_k)(1-f_l)\right] \nonumber\\
    & &+ \sum\limits_{k,l}\langle k | j_i | l\rangle\langle\bar{k} | j_j |
        \bar{l}\rangle \times u_k v_k u_l v_l \left[-
        e^{\tau(E_k-E_l)}f_k(1-f_l)-e^{-\tau(E_k-E_l)}(1-f_k)f_l\right .\\
    & &  \hspace{4.5 cm} + \left .
        e^{\tau(E_k+E_l)}f_k f_l+e^{-\tau(E_k+E_l)}(1-f_k)(1-f_l)\right] \;.
\end{eqnarray}
The sum in the first term is zero by symmetry (we assume axially symmetric
deformed states). To proceed, we need to carefully examine the behavior
of the matrix elements of $j_i$ under time reversal. Denoting by ${\cal
T}$ the time reversal operator (cf. Ref.~\cite{bm69}, 1-2c), we have
${\cal T}j_i{\cal T}^{-1}= -j_i$, and for $k>0$ (i.e., for spin
projection $\mu>0$)
\be
\begin{array}{l}{\cal T}| k\rangle={\cal T} | q,\;\mu\rangle=|q,\;-\mu
\rangle  = | {\bar k}\rangle \\
{\cal T}| -k\rangle={\cal T} | q,\;-\mu\rangle=-|q,\;\mu\rangle = -|k\rangle \;.
\end{array}
\ee
Using the transformation properties of the matrix elements under time
reversal [see (1-34) in \cite{bm69}], we then have
\be\label{me-time}
\begin{array}{l}
\langle k | j_i | l\rangle=-\langle\bar{l} | j_i | \bar{k}\rangle\;\;{\rm
for}\;\;
k\cdot l>0 \\
\langle k | j_i | l\rangle=\langle\bar{l} | j_i | \bar{k}\rangle\;\; {\rm
for}\;\;
k\cdot l<0 \;.\end{array}
\ee
Using relations (\ref{me-time}), we can rewrite (\ref{j-j-2}) in the form
\begin{eqnarray}\label{j-j-3}
    \langle J_i(\tau)J_j(0)\rangle = &
    \sum\limits_{k,l>0}\left(\langle k | j_i | l\rangle
        \langle l | j_j | k\rangle+\langle l | j_i | k\rangle
        \langle k | j_j | l\rangle+\langle k | j_i | -l\rangle
        \langle -l | j_j | k\rangle+\langle -k | j_i | l\rangle
        \langle l | j_j | -k\rangle\right)\\
    & \times \left[e^{\tau(E_k-E_l)}(u_k^2 u_l^2+u_k v_k u_l v_l)f_k(1-f_l)
        +e^{-\tau(E_k-E_l)}(v_k^2 v_l^2+u_k v_k u_l v_l)(1-f_k)f_l
        \right .\\
    & \left . +
        e^{\tau(E_k+E_l)}(u_k^2 v_l^2-u_k v_k u_l v_l)f_k f_l+
        e^{-\tau(E_k+E_l)}(v_k^2 u_l^2-u_k v_k u_l v_l)(1-f_k)(1-f_l)\right] \;.
\end{eqnarray}
The intrinsic moments are given by
${\cal I}_{ij}=\int_0^\beta d\tau\,\langle J_i(\tau)J_j(0)\rangle$. Using
(\ref{j-j-3}) and the relations
\begin{equation}\begin{array}{l}
{\displaystyle f_k(1-f_l)\int_0^\beta e^{\tau(E_k-E_l)}d\tau=(1-f_k)f_l
\int_0^\beta e^{-\tau(E_k-E_l)}d\tau={f_l-f_k\over E_k-E_l}}\vspace{.2cm}\\
{\displaystyle f_k f_l\int_0^\beta e^{-\tau(E_k+E_l)}d\tau=(1-f_k)(1-f_l)
\int_0^\beta e^{\tau(E_k+E_l)}d\tau={1-f_k-f_l\over E_k+E_l}}\;,
\end{array}\end{equation}
we arrive at Eq.~(\ref{inertia-qp}).

\section*{Appendix B: derivation of the effective moment of inertia}

  Here we derive Eq.~(\ref{inertia3}). The components $L_\mu$ of the
   angular momentum in the laboratory frame are related to the
   components $J_i$ in the intrinsic
frame through $L_\mu=\sum_{\mu^{'}} {{\cal D}_{\mu
\mu'}^{(1)}}^*(\psi,\theta,\phi)J_{\mu^{'}}$, where ${\cal D}^{(1)}$
are Wigner rotation matrices and $(\psi,\theta,\phi)$ are Euler angles.
In particular
$L_z=\sum_i \hat\omega_i J_i$ where the unit vector $\hat \omega$ is given by
$\hat{\omega}=(-\sin \theta \cos \phi, \sin \theta \sin \phi, \cos \theta)$.
To restore rotational invariance, we integrate
over all the possible orientations of the intrinsic frame. In practice we
replace in (\ref{free1})
\be\label{restore}
e^{-\beta(H'-\omega L_z)}\rightarrow {1\over 4\pi} \int_0^{2\pi} d\phi\;
\int_0^\pi d\theta\; \sin \theta e^{-\beta(H'-\omega \sum_i
\hat{\omega_i} J_i)} \;,
\ee
and use (\ref{inertia1}) to find
\begin{equation}\label{inertia2}
I = {1\over 4\pi} \int_0^{2\pi} d\phi\;
\int_0^\pi d\theta\; \sin \theta
\sum_{ij} \hat\omega_i I_{ij} \hat\omega_j \;.
\end{equation}
where $I_{ij}$ is the intrinsic moment of inertia tensor
(\ref{inertia-tensor}). The explicit integration of (\ref{inertia2}) leads to
(\ref{inertia3}).
\end{widetext}

\end{document}